\documentclass{article}
\usepackage{arxiv}       
\usepackage{float}
\usepackage{placeins}           

\usepackage[utf8]{inputenc}     
\usepackage[T1]{fontenc}        
\usepackage{hyperref}           
\usepackage{url}                

\hypersetup{
    colorlinks = true,
    linkcolor  = black,                 
    citecolor  = {[rgb]{0.0,0.2,0.6}},   
    urlcolor   = {[rgb]{0.0,0.2,0.6}},   
    pdfborder  = {0 0 0}
}
\usepackage{booktabs}           
\usepackage{amsmath}            
\usepackage{amssymb}            
\usepackage{graphicx}           
\usepackage{multirow}           
\usepackage{xcolor}             
\usepackage{listings}           
\usepackage{microtype}          
\usepackage[numbers]{natbib}    

\title{Where Do Large Language Models Fail on Competitive Programming?\\
       A Taxonomy of Failures by Algorithm Type and Difficulty Rating}

\author{
  \textbf{Ayush Kumar Jha} \\
  Department of Information Technology \\
  IIIT Bhubaneswar \\
  Bhubaneswar, Odisha, India \\
  \texttt{ayushjha4277@gmail.com}
  \and
  \textbf{Shalini Jha} \\
  Department of Computer Science \\
  IIT Patna \\
  Patna, Bihar, India \\
  \texttt{jha718897@gmail.com}
}
\date{\today}

\renewcommand{\headeright}{Preprint}
\renewcommand{\undertitle}{Preprint — Under Review}
\renewcommand{\shorttitle}{LLM Failure Taxonomy on Competitive Programming}

\begin{document}
\maketitle

\begin{abstract}
Large language models (LLMs) demonstrate increasing proficiency on competitive programming benchmarks, yet technical reports predominantly publish aggregate pass rates, obscuring domain-specific vulnerabilities. We present a systematic empirical study of LLM failure patterns using a balanced taxonomy of 315 Codeforces problems across seven algorithm categories and three difficulty tiers. We evaluate GPT-4o and Claude Sonnet 4.6 under strict execution-based conditions, controlling for temperature ($T=0.2$). To isolate the impact of reasoning frameworks on algorithmic correctness, we conduct an ablation study comparing direct zero-shot generation against zero-shot Chain-of-Thought (CoT). Our findings reveal a severe divergence from standard NLP benchmarks: forcing CoT aggressively penalizes GPT-4o, dropping its pass rate from 46.0\% to 36.8\% and exacerbating a critical weakness in Greedy logic. Conversely, while Claude maintains a higher logical baseline (63.5\% under CoT), the expanded text generation severely degrades its markdown instruction adherence, causing its Compile Errors to more than triple (from 9 to 31, a 244\% increase). Furthermore, failure-mode analysis indicates that Wrong Answer (WA) is the dominant verdict for both models---accounting for over 90\% of GPT-4o's and roughly 70\% of Claude's unaccepted solutions. These findings empirically demonstrate that standard prompt engineering techniques fail to bridge the algorithmic reasoning gap in competitive programming environments.
\end{abstract}

\keywords{large language models \and code generation \and
          competitive programming \and failure analysis \and
          algorithm taxonomy \and GPT-4o \and Claude}

\section{Introduction}
\label{sec:intro}
Large Language Models (LLMs) have demonstrated substantial capability in automated code generation. Early benchmarks, such as HumanEval \citep{chen2021codex}, evaluated basic functional synthesis, while subsequent datasets like APPS \citep{hendrycks2021apps} and CodeContests \citep{deepmind2022codecontests} shifted the evaluation frontier to competitive programming. Competitive programming requires models to parse complex natural language constraints, identify optimal algorithmic substructures, and synthesize executable code that strictly adheres to asymptotic time and memory limits.

While frontier models including AlphaCode \citep{li2022competition}, GPT-4 \citep{openai2023gpt4}, and the Claude 3 family \citep{anthropic2024claude} have achieved increasingly high aggregate pass rates on these benchmarks, the monolithic reporting of these metrics obscures the specific boundaries of model capabilities. Current technical reports provide percentile rankings or overall zero-shot accuracy, but fail to provide granular taxonomies of \textit{where} and \textit{how} these models fail across distinct algorithmic domains.

We address this gap by presenting a systematic empirical study of LLM failure patterns. Using a highly controlled, balanced sample of 315 Codeforces problems, we map the performance of two high-efficiency frontier models (GPT-4o and Claude Sonnet 4.6) across seven distinct algorithm categories and three difficulty tiers. 

\subsection{Motivation}
\label{sec:intro:motivation}
Understanding the specific algorithmic blind spots of LLMs is necessary for the development of targeted self-correction pipelines and specialized code-generation agents. If a model fails predominantly on syntax or compilation, the intervention requires better linters or compiler-feedback loops. However, if a model reliably compiles but fails on mathematical logic or edge-case handling, the intervention requires algorithmic reasoning frameworks. By categorizing failures simultaneously by problem type, difficulty, and execution verdict, we provide practitioners and benchmark designers with a diagnostic map of current model limitations.

\subsection{Research Questions}
\label{sec:intro:rq}
Our empirical evaluation is structured around three primary research questions:
\begin{itemize}
    \item[\textbf{RQ1}] Does LLM performance degrade non-uniformly across specific algorithm categories (e.g., Dynamic Programming, Graphs, Greedy logic)?
    \item[\textbf{RQ2}] How does the empirical pass rate correlate with official Codeforces difficulty ratings, and where do frontier models cross the threshold of failure?
    \item[\textbf{RQ3}] What is the primary execution failure mode (Wrong Answer, Time Limit Exceeded, Runtime Error, or Compile Error) when models fail to generate an accepted solution?
\end{itemize}

\subsection{Contributions}
\label{sec:intro:contributions}
Our methodology and findings yield the following contributions to the evaluation of code-generation models:
\begin{enumerate}
    \item \textbf{A 2D Failure Taxonomy:} We introduce a novel evaluation grid isolating LLM performance across both Algorithm Category and Difficulty Tier, proving that pass rates degrade non-linearly depending on the required algorithmic domain.
    \item \textbf{The CoT Penalty in Code Generation:} Through a controlled ablation study, we demonstrate that zero-shot Chain-of-Thought prompting—a standard technique for improving LLM reasoning—catastrophically degrades GPT-4o's performance ($-9.2$ percentage points in AC rate) while destroying Claude Sonnet 4.6's formatting adherence. 
    \item \textbf{Execution Error Profiling:} We demonstrate that the majority of unaccepted solutions result from a Wrong Answer (WA) verdict (over 90\% for GPT-4o), indicating the bottleneck is fundamentally algorithmic reasoning rather than syntactical validity or asymptotic efficiency.
\end{enumerate}
\section{Related Work}
\label{sec:related}
Our evaluation intersects with three primary subfields of machine learning research: foundational code generation benchmarks, frontier model evaluations, and the analysis of language model failure modes.

\subsection{Foundational Code Generation Benchmarks}
\label{sec:related:codegen}
Initial evaluations of LLM coding capabilities relied on zero-shot generation of isolated functions. HumanEval \citep{chen2021codex} established a baseline for measuring syntactical correctness and basic functional logic using a pass@$k$ metric. To evaluate deeper algorithmic reasoning, \citet{hendrycks2021apps} introduced the APPS dataset, shifting the focus to competitive programming problems sourced from platforms like Codeforces and LeetCode. While these benchmarks defined the standard for evaluating code generation, they predominantly report aggregate pass rates. Our work extends this methodology by disaggregating performance, providing a fine-grained taxonomy of where these aggregate metrics fail across specific algorithm categories and difficulty tiers.

\subsection{Frontier Models and Competitive Programming}
\label{sec:related:cp}
Competitive programming serves as a rigorous testbed for LLM reasoning capabilities due to its strict execution constraints and requirement for multi-step logical synthesis. AlphaCode \citep{li2022competition} demonstrated that transformer-based architectures could achieve median human-level performance in simulated programming contests using massive-scale sampling and filtering. Subsequent open-weight models, such as DeepSeek-Coder \citep{guo2024deepseek}, and proprietary frontier models like GPT-4 \citep{openai2023gpt4} and the Claude 3 family \citep{anthropic2024claude}, have further optimized for competitive programming tasks. However, the technical reports accompanying these models typically publish monolithic success metrics (e.g., simulated Elo ratings or aggregate pass@1 scores). These reports obscure the models' specific algorithmic blind spots, which our 2D taxonomy matrix explicitly maps.

\subsection{Evaluation Contamination}
\label{sec:related:contamination}
A known confounding variable in competitive programming evaluations is dataset contamination. \citet{jain2025livecodebench} introduced LiveCodeBench, demonstrating that frontier LLMs achieve inflated performance on historical LeetCode and Codeforces problems due to memorization of their pre-training corpora. By establishing a contamination-free evaluation window, they proved that true zero-shot reasoning degrades significantly on novel problems. We acknowledge that our CodeContests \citep{deepmind2022codecontests} evaluation set predates the training cutoffs for both GPT-4o and Claude Sonnet 4.6. Consequently, our study does not claim to measure uncontaminated zero-shot reasoning; rather, it measures the models' combined retrieval and reasoning capabilities. Given this baseline, the severe degradation we observe on specific algorithmic subsets (such as hard dynamic programming and greedy logic) represents a fundamental failure in the models' ability to synthesize learned concepts, regardless of prior exposure.

\subsection{Failure Analysis and Self-Correction}
\label{sec:related:failure}
Recent literature investigates how LLMs resolve generation failures through iterative prompting. Techniques such as Chain-of-Thought (CoT) \citep{wei2022chain} structure the reasoning process, while frameworks like Reflexion \citep{shinn2023reflexion} and Self-Refine \citep{madaan2023selfrefine} attempt to correct code using verbal reinforcement and compiler feedback. Despite these advancements, \citet{olausson2024selfrepair} empirically demonstrated that LLMs frequently fail to self-repair logical bugs without external human guidance. Their findings indicate that while models easily resolve syntax errors, they struggle to identify and correct flawed algorithmic logic. This limitation directly motivates our experimental design; by restricting evaluation to a single-attempt (pass@1) zero-shot prompt, we isolate the models' base logical pathways. Our finding that failures are overwhelmingly dominated by Wrong Answer (WA) rather than Compile Error (CE) verdicts provides empirical reinforcement for Olausson et al.'s observations within a strictly categorized competitive programming taxonomy.

\section{Methodology}
\label{sec:method}
To systematically evaluate where Large Language Models fail on competitive programming tasks, we designed an automated, zero-shot evaluation pipeline. This pipeline samples problems across algorithmic domains, queries models under controlled hyperparameters, executes the generated Python code in an isolated sandbox, and classifies failures into a standard competitive programming verdict hierarchy.

\subsection{Dataset}
\label{sec:method:dataset}
We construct our evaluation benchmark using the CodeContests dataset \citep{deepmind2022codecontests}, originally introduced to evaluate AlphaCode \citep{li2022competition}. While CodeContests aggregates problem statements from multiple competitive programming platforms, we filter the dataset to exclusively include problems sourced from Codeforces. This filtering is strictly necessary to access high-quality, community-verified metadata—specifically, the \texttt{cf\_tags} (algorithm categorisations) and \texttt{cf\_rating} (Elo-based difficulty scores) required to build our taxonomy. 

\subsection{Problem Categorisation}
\label{sec:method:categories}
To map model performance across distinct logical domains, we group the official Codeforces problem tags into seven broad algorithm categories. Problems containing multiple tags are assigned to the first matching category. The taxonomy mapping is detailed in Table \ref{tab:categories}.

\begin{table}[h]
  \centering
  \caption{Algorithm categories and their corresponding Codeforces tags.}
  \label{tab:categories}
  \begin{tabular}{ll}
    \toprule
    \textbf{Category} & \textbf{Codeforces Tags} \\
    \midrule
    Dynamic Programming   & \texttt{dp} \\
    Graphs                & \texttt{graphs, dfs and similar, shortest paths, trees} \\
    Greedy                & \texttt{greedy} \\
    Binary Search         & \texttt{binary search, two pointers} \\
    Mathematics           & \texttt{math, number theory, combinatorics} \\
    Data Structures       & \texttt{data structures, sortings} \\
    Implementation        & \texttt{implementation, brute force} \\
    \bottomrule
  \end{tabular}
\end{table}

\subsection{Difficulty Tiers}
\label{sec:method:difficulty}
We stratify the problems into three difficulty tiers based on their official Codeforces Elo problem ratings. These cutoffs are not arbitrary; they align directly with Codeforces' official user rank thresholds and Division boundaries:
\begin{itemize}
    \item \textbf{Easy (800--1400):} Corresponds to the Newbie and Pupil ranks. Problems typically require introductory logic and basic implementation (standard Division 3/4 difficulty).
    \item \textbf{Medium (1401--1900):} Corresponds to the Specialist and Expert ranks. Problems require intermediate algorithms and standard data structures (standard Division 2 difficulty).
    \item \textbf{Hard (1901--3500):} Corresponds to Candidate Master and above. Problems require advanced multi-step optimization and complex mathematics (standard Division 1 difficulty).
\end{itemize}

\subsection{Sampling Strategy}
\label{sec:method:sampling}
To prevent algorithmic imbalance from skewing the aggregate pass rates, we apply a strict balanced sampling strategy. We extract exactly 15 problems for each intersection of Algorithm Category and Difficulty Tier (7 categories $\times$ 3 tiers $\times$ 15 problems), yielding a highly controlled benchmark of 315 problems. To ensure reproducibility, problems are selected deterministically based on their original index in the CodeContests training split.

\subsection{Models Evaluated}
\label{sec:method:models}
We evaluate two widely deployed, high-efficiency frontier models: GPT-4o \citep{openai2024gpt4o} and Claude Sonnet 4.6 (released February 2026) \citep{anthropic2026sonnet46}. To minimize output variance and measure the models' most confident logical pathways, we set the sampling temperature to $T=0.2$. The maximum output length is constrained to 2048 tokens to accommodate verbose reasoning and complex string manipulations without truncation. We evaluate using a strict \textit{pass@1} metric; each model is given only a single generation attempt per problem.

\subsection{Prompt Design and Ablation Setup}
\label{sec:method:prompt}
To isolate the impact of reasoning frameworks on algorithmic accuracy, both models are evaluated under two strict, zero-shot prompting conditions. We structure the input using a standard user prompt that dynamically injects the problem metadata, paired with a condition-specific system prompt. 

The universal user prompt applied across all evaluations is formatted as follows:
\begin{lstlisting}[basicstyle=\small\ttfamily, breaklines=true]
Problem: {problem['name']}
{problem['description']}

Write a complete Python 3 solution.
\end{lstlisting}

\textbf{Condition A: Direct Generation.} This baseline strictly forbids natural language explanations, forcing the model to rely on its implicit internal representations and maximizing the token budget for code generation.
\begin{lstlisting}[basicstyle=\small\ttfamily, breaklines=true]
You are a competitive programmer. Solve the given problem in Python 3.
Output ONLY the complete Python code inside a markdown block. No explanations.
The code must read from stdin and print to stdout.
\end{lstlisting}

\textbf{Condition B: Chain-of-Thought (CoT).} This condition overrides the Direct baseline by forcing explicit algorithmic planning prior to code generation. This tests whether enforced exchange-arguments and formal mathematical derivations resolve the models' logical blind spots.
\begin{lstlisting}[basicstyle=\small\ttfamily, breaklines=true]
You are a competitive programmer. Solve the given problem in Python 3.
First, write a detailed algorithmic plan and mathematical proof of your approach.
Then, output the complete Python code inside a markdown block.
The code must read from stdin and print to stdout.
\end{lstlisting}
\subsection{Evaluation Protocol}
\label{sec:method:eval}
Model outputs are parsed via regular expressions to extract the executable Python code, which is then run inside a \texttt{subprocess} sandbox. Each solution is executed against a maximum of three public test cases. To ensure a fair assessment of algorithmic efficiency, we enforce a strict 5-second timeout limit per test case. 

Results are classified into a standard competitive programming verdict hierarchy. If a solution fails multiple test cases, the worst verdict is recorded based on the following priority list:
\begin{enumerate}
    \item \textbf{Compile Error (CE):} Invalid Python syntax.
    \item \textbf{Runtime Error (RE):} Exceptions during execution (e.g., \texttt{IndexError}, \texttt{ZeroDivisionError}).
    \item \textbf{Time Limit Exceeded (TLE):} Execution exceeds the 5-second threshold, indicating suboptimal asymptotic complexity.
    \item \textbf{Wrong Answer (WA):} The program compiles and executes within the time limit, but produces incorrect output.
    \item \textbf{Accepted (AC):} The program produces the exact expected output for all test cases.
\end{enumerate}

\section{Results}
\label{sec:results}
In this section, we present the empirical pass rates and failure distributions across the 315 evaluated problems. We structure our findings to systematically address the core research questions: aggregate performance, degradation by algorithm and difficulty, and the distribution of execution failure modes. Unless stated otherwise, all per-category, per-tier, joint, and failure-mode results report the Chain-of-Thought (CoT) condition; the Direct condition is reported only as the ablation baseline in Section~\ref{sec:results:overall}.

\subsection{Overall Performance}
\label{sec:results:overall}
Across the strictly controlled, balanced sample of 315 problems, Claude Sonnet 4.6 established a higher empirical baseline than GPT-4o under both prompting conditions. Under \textit{Direct} generation, Claude achieved an Accepted (AC) verdict on 196 problems (62.2\%) versus GPT-4o's 145 (46.0\%). Under \textit{Chain-of-Thought} prompting, Claude solved 200 problems (63.5\%) while GPT-4o dropped to 116 (36.8\%). Claude thus remained essentially stable across the two conditions, whereas enforced CoT degraded GPT-4o by $9.2$ percentage points---a divergence we analyze in detail in Section~\ref{sec:discussion}.

\subsection{Performance by Algorithm Category (RQ1)}
\label{sec:results:algo}
Model performance exhibits high variance depending on the underlying algorithmic domain required to solve the problem. Table \ref{tab:algo_pass_rates} details the pass rates across all seven categories under CoT prompting.

\begin{figure}[htbp]
    \centering
    \includegraphics[width=0.85\linewidth]{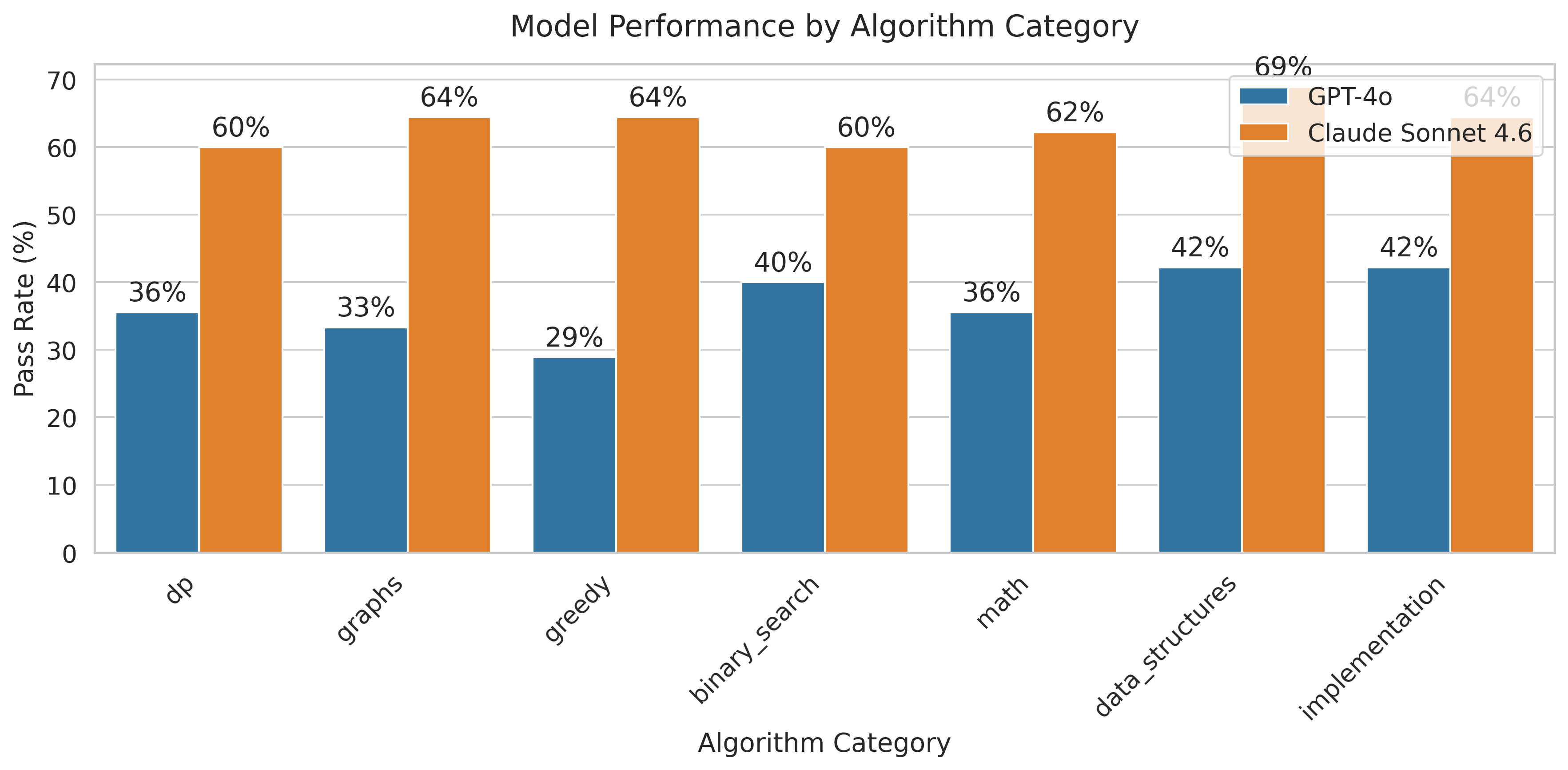}
    \caption{Aggregate pass rates by algorithm category under CoT prompting. GPT-4o's vulnerability to "context poisoning" is most evident in the Greedy category, where its pass rate degraded to 29\%.}
    \label{fig:algo_bars}
\end{figure}

Claude Sonnet 4.6 outperformed GPT-4o in every category. The performance gap is most pronounced in Greedy and Graphs logic. On \texttt{greedy} problems, Claude reached 64.4\% versus GPT-4o's 28.9\%---a gap of 35.5 percentage points (Figure~\ref{fig:algo_bars}). Greedy was also GPT-4o's single weakest category (28.9\%), consistent with the ``context-poisoning'' effect we analyze in Section~\ref{sec:discussion:greedy}.

\begin{table}[htbp]
  \centering
  \caption{Pass rate (\%) by algorithm category under CoT prompting ($n=45$ per category).}
  \label{tab:algo_pass_rates}
  \begin{tabular}{lcc}
    \toprule
    \textbf{Algorithm Category} & \textbf{GPT-4o} & \textbf{Claude Sonnet 4.6} \\
    \midrule
    Dynamic Programming & 35.6 & 60.0 \\
    Graphs              & 33.3 & 64.4 \\
    Greedy              & 28.9 & 64.4 \\
    Binary Search       & 40.0 & 60.0 \\
    Mathematics         & 35.6 & 62.2 \\
    Data Structures     & 42.2 & 68.9 \\
    Implementation      & 42.2 & 64.4 \\
    \bottomrule
  \end{tabular}
\end{table}

\subsection{Performance by Difficulty Tier (RQ2)}
\label{sec:results:difficulty}
We observe a non-linear degradation in pass rates as problem difficulty scales. As shown in Table \ref{tab:diff_pass_rates}, both models reliably solve introductory logic, with Claude and GPT-4o achieving 91.4\% and 61.9\% on Easy problems, respectively. 

However, scaling to Medium difficulty exposes a sharp divergence in model capabilities. GPT-4o experiences a steep 22.9 percentage point drop in accuracy (falling to 39.0\%), whereas Claude demonstrates higher robustness, retaining a 70.5\% pass rate on intermediate data structures and algorithms. Both models collapse on Hard problems, falling to 28.6\% (Claude) and 9.5\% (GPT-4o)---below the 30\% threshold.

\begin{table}[htbp]
  \centering
  \caption{Pass rate (\%) by difficulty tier under CoT prompting ($n=105$ per tier).}
  \label{tab:diff_pass_rates}
  \begin{tabular}{lcc}
    \toprule
    \textbf{Difficulty Tier} & \textbf{GPT-4o} & \textbf{Claude Sonnet 4.6} \\
    \midrule
    Easy (800--1400)   & 61.9 & 91.4 \\
    Medium (1401--1900)& 39.0 & 70.5 \\
    Hard (1901--3500)  & 9.5  & 28.6 \\
    \bottomrule
  \end{tabular}
\end{table}

\subsection{Joint Analysis: Algorithm $\times$ Difficulty}
\label{sec:results:grid}
\begin{figure}[htbp]
    \centering
    \includegraphics[width=\textwidth]{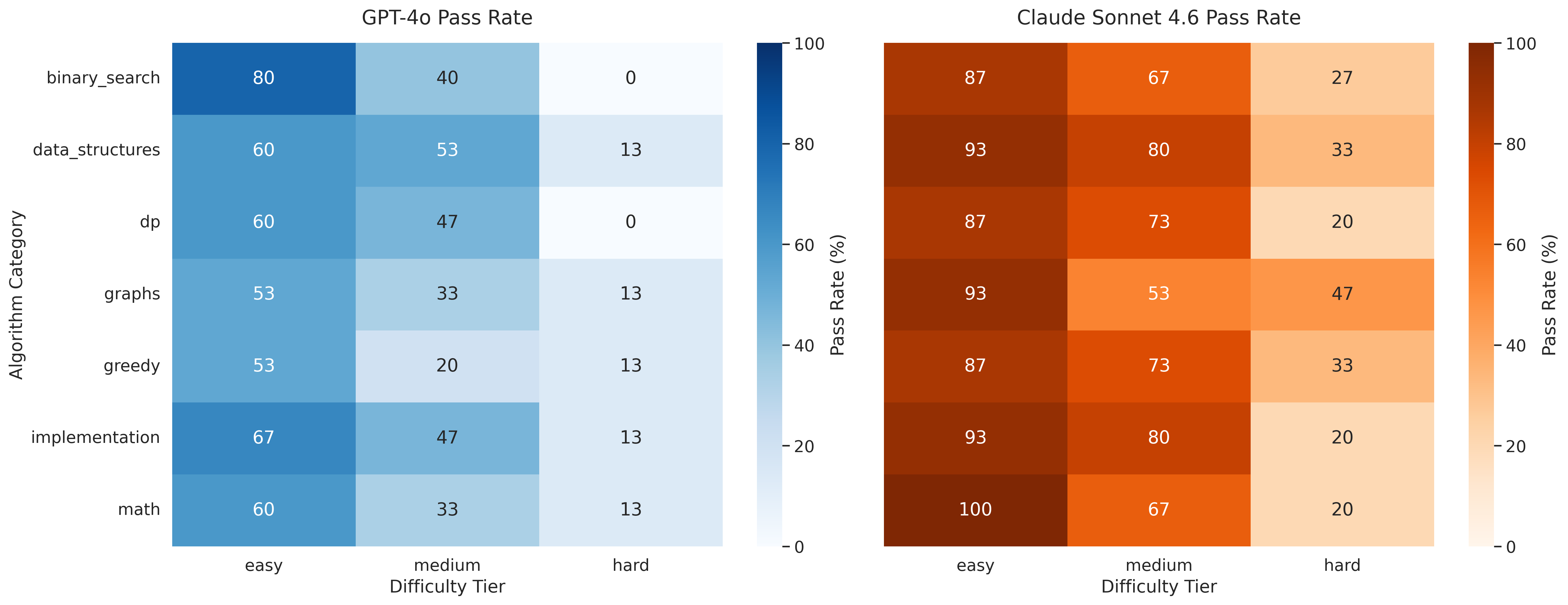}
    \caption{Pass rates (\%) across the Algorithm $\times$ Difficulty taxonomy under the Chain-of-Thought (CoT) condition. GPT-4o exhibits a severe capability collapse at the Hard tier (dropping to 0\% on DP and Binary Search), while Claude Sonnet 4.6 maintains structural stability.}
    \label{fig:heatmap}
\end{figure}

To isolate specific algorithmic bottlenecks, we expand the evaluation into a 2D taxonomy matrix (Table \ref{tab:grid}), reporting the CoT condition visualized in Figure \ref{fig:heatmap}. This granular breakdown reveals domain-specific failure thresholds. 

Dynamic Programming (DP) proves universally brittle for LLMs at higher difficulties; on Hard DP problems, GPT-4o solves none (0\%) while Claude manages only 20\%. GPT-4o's collapse at the Hard tier is most severe in DP and Binary Search, where it fails every problem (0\%). Claude outperforms GPT-4o in every cell of the matrix; its largest Hard-tier advantage appears in Graphs (47\% vs.\ 13\%), where it identifies structural graph-theoretic templates that GPT-4o misses. We note that the Direct baseline produces a different picture at the extremes—most notably an inversion on Hard Mathematics in which GPT-4o briefly overtakes Claude—which we analyze in Section~\ref{sec:discussion:math}.

\begin{table*}[h]
  \centering
  \caption{Taxonomy of Pass Rates (\%) across Algorithm Category $\times$ Difficulty Tier, under CoT prompting. Format: GPT-4o / Claude Sonnet 4.6. Bold text denotes the superior pass rate for the given cell.}
  \label{tab:grid}
  \begin{tabular}{lccc}
    \toprule
    \textbf{Algorithm} & \textbf{Easy} & \textbf{Medium} & \textbf{Hard} \\
    \midrule
    Dynamic Programming & 60 / \textbf{87} & 47 / \textbf{73} &  0 / \textbf{20} \\
    Graphs              & 53 / \textbf{93} & 33 / \textbf{53} & 13 / \textbf{47} \\
    Greedy              & 53 / \textbf{87} & 20 / \textbf{73} & 13 / \textbf{33} \\
    Binary Search       & 80 / \textbf{87} & 40 / \textbf{67} &  0 / \textbf{27} \\
    Mathematics         & 60 / \textbf{100} & 33 / \textbf{67} & 13 / \textbf{20} \\
    Data Structures     & 60 / \textbf{93} & 53 / \textbf{80} & 13 / \textbf{33} \\
    Implementation      & 67 / \textbf{93} & 47 / \textbf{80} & 13 / \textbf{20} \\
    \bottomrule
  \end{tabular}
\end{table*}

\subsection{Failure Mode Analysis (RQ3)}
\label{sec:results:failures}
Evaluating the raw distribution of verdict labels provides crucial insight into the nature of LLM coding failures. For solutions that did not achieve an AC verdict under CoT prompting, we isolate the specific execution bottlenecks (Table \ref{tab:failure_modes}).

\begin{figure}[htbp]
    \centering
    \includegraphics[width=0.85\linewidth]{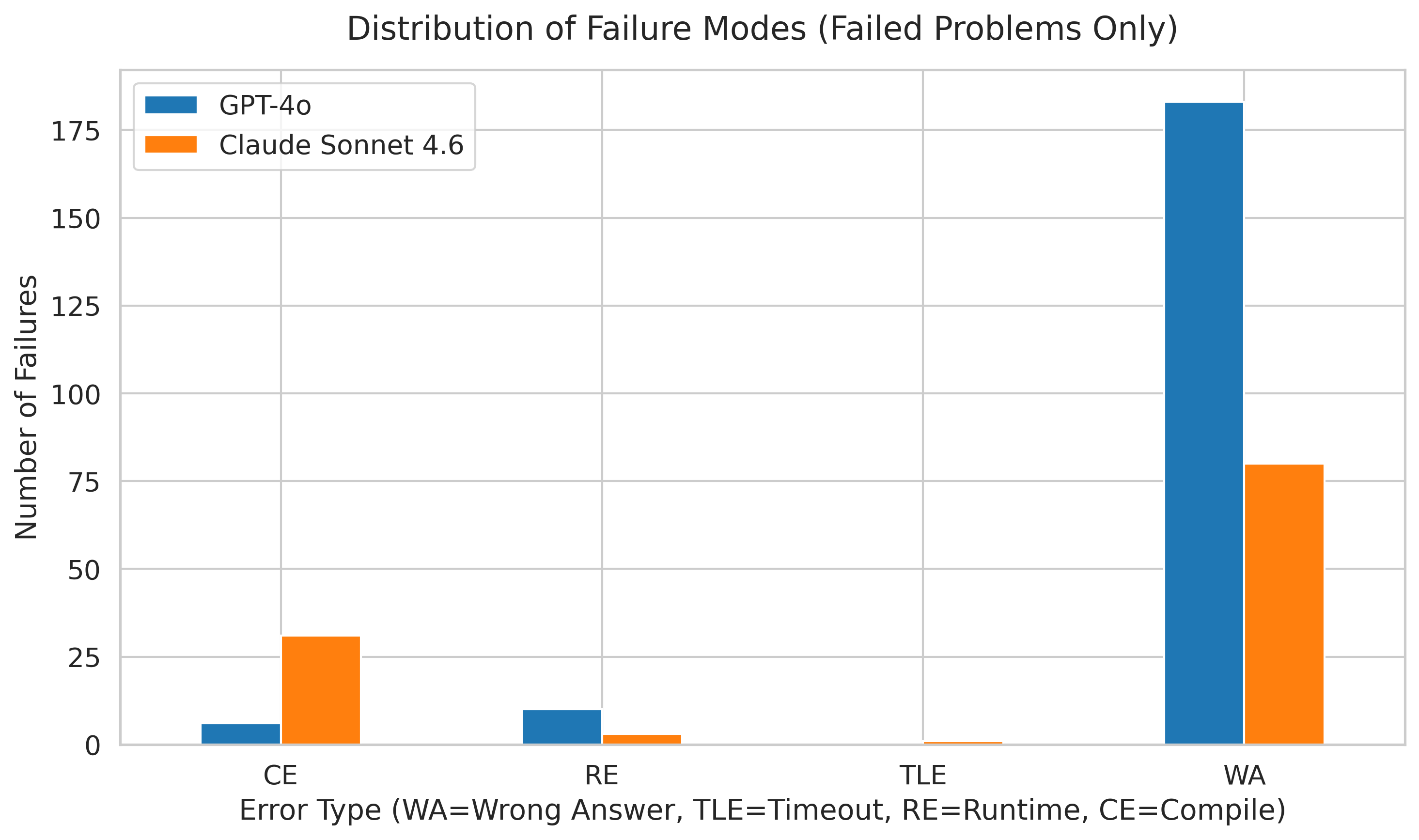}
    \caption{Distribution of failure verdicts under CoT prompting. The requirement to generate verbose mathematical proofs induced a formatting collapse in Claude Sonnet 4.6, resulting in a spike of Compile Errors (CE), whereas GPT-4o's failures remain overwhelmingly logical (WA).}
    \label{fig:failure_modes}
\end{figure}

The data demonstrates that failures are overwhelmingly logical rather than syntactical or efficiency-based (Figure~\ref{fig:failure_modes}). Wrong Answer (WA) accounts for 92.0\% (183/199) of GPT-4o's failures and 69.6\% (80/115) of Claude's failures. This indicates that both models reliably generate executable Python code that runs within the time limit (evidenced by only a single cumulative Time Limit Exceeded failure), but fail to correctly implement the underlying mathematical proofs or handle edge cases. 

\begin{figure}[htbp]
    \centering
    \includegraphics[width=0.85\linewidth]{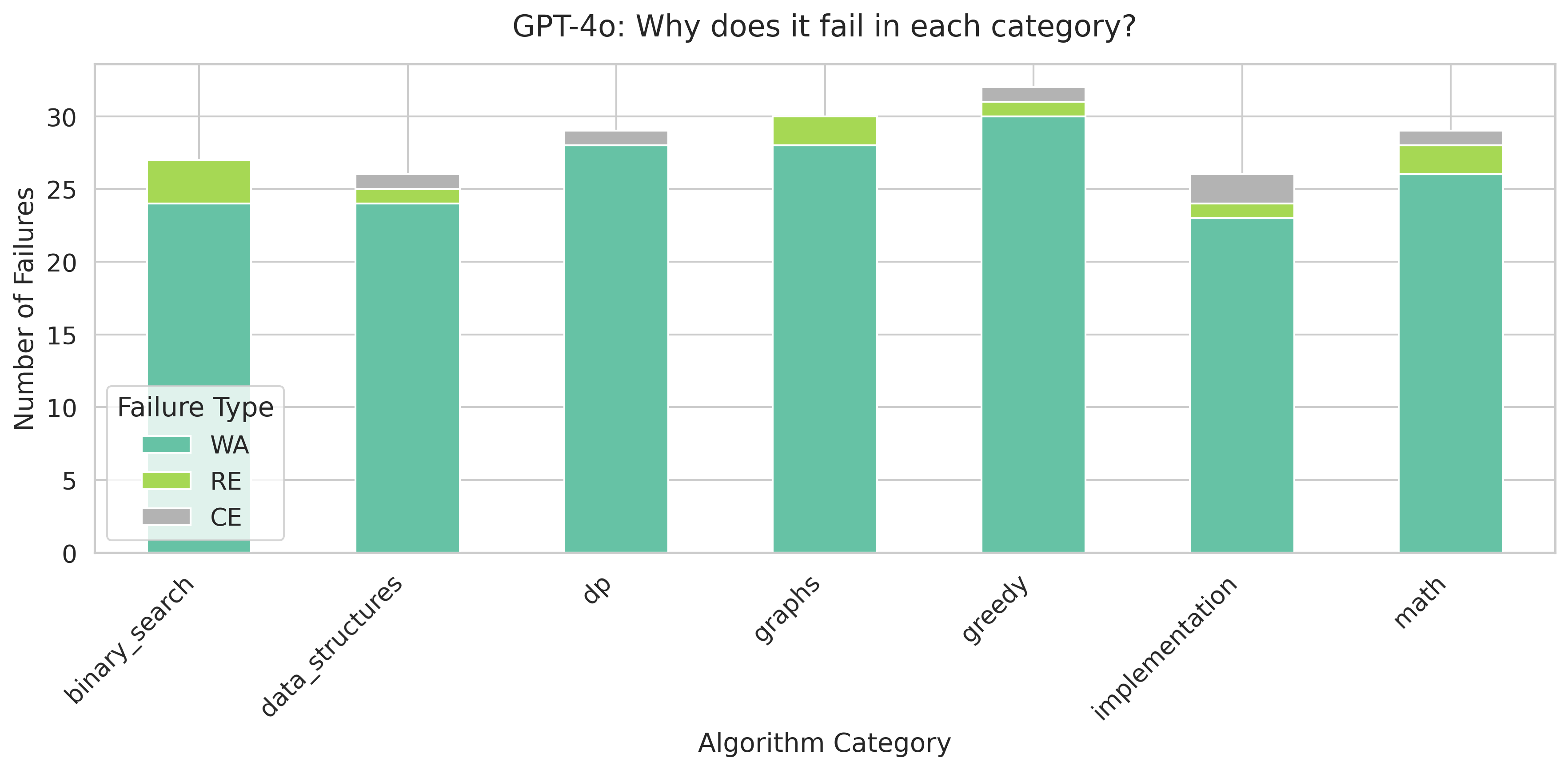}
    \caption{Distribution of GPT-4o's failure types across algorithm categories. The uniform dominance of Wrong Answer (WA) confirms that the model consistently generates syntactically valid code that fails implicit correctness proofs.}
    \label{fig:gpt4o_errors}
\end{figure}

The two models diverge in their secondary failure modes. Claude's failures include a pronounced spike in Compile Errors (31 CE)—consistent with the formatting collapse induced by verbose CoT generation (Section~\ref{sec:discussion:math})—alongside only 3 Runtime Errors (RE). GPT-4o, by contrast, produced just 6 CE but 10 RE. Broken down by algorithm category, GPT-4o's failures are uniformly dominated by WA across every domain (Figure~\ref{fig:gpt4o_errors}), reinforcing that its bottleneck is correctness rather than syntax.

\begin{table}[htbp]
  \centering
  \caption{Distribution of failure modes for unaccepted solutions under CoT prompting.}
  \label{tab:failure_modes}
  \begin{tabular}{lcc}
    \toprule
    \textbf{Verdict} & \textbf{GPT-4o (Failures: 199)} & \textbf{Claude (Failures: 115)} \\
    \midrule
    Wrong Answer (WA)         & 183 & 80 \\
    Compile Error (CE)        & 6   & 31 \\
    Runtime Error (RE)        & 10  & 3  \\
    Time Limit Exceeded (TLE) & 0   & 1  \\
    \bottomrule
  \end{tabular}
\end{table}

\FloatBarrier
\section{Discussion}
\label{sec:discussion}
The empirical results demonstrate that LLM failure on competitive programming tasks is highly sensitive to the underlying algorithmic paradigm. In this section, we provide a qualitative failure analysis of the anomalous phenomena observed in the 2D taxonomy matrix, leveraging code forensics from both the Direct and CoT evaluations to distinguish between reasoning failures and execution bottlenecks.

\subsection{The CoT Penalty and Context Poisoning in Greedy Algorithms}
\label{sec:discussion:greedy}
The most counter-intuitive finding in our dataset is the catastrophic degradation of GPT-4o under CoT prompting (dropping from a 46.0\% to 36.8\% aggregate AC rate). This deficit is most pronounced in the \texttt{greedy} category, where performance dropped to 28.9\%.

Code forensics reveal a phenomenon we characterize as "context poisoning" driven by proof obligation blindness. Greedy algorithms require an implicit correctness proof—typically an exchange argument—to guarantee that a local choice yields a global optimum. When forced by the CoT prompt to explicitly write out this algorithmic plan, GPT-4o frequently hallucinates a provably incorrect heuristic. Because self-attention mechanisms heavily weight the immediate preceding context, this flawed natural language proof appears to bind the subsequent Python generation to the incorrect logic. GPT-4o performs significantly better when allowed to bypass explicit proof generation and rely on its implicit coding intuition.

\subsection{Formatting Collapse and the Math Inversion}
\label{sec:discussion:math}
While Claude Sonnet 4.6 did not suffer a logical penalty under CoT, it suffered a mechanical one. The requirement to generate detailed proofs caused Claude to output massive text blocks (frequently 1,000+ tokens) before initializing its Python markdown environment. Even at a controlled temperature of $T=0.2$, this extended generation severely degraded its instruction adherence, causing Compile Errors to more than triple, rising from 9 in the Direct run to 31 in the CoT run.

The Direct baseline additionally exhibited an anomalous Hard Mathematics inversion. Under Direct generation, GPT-4o beat Claude on Hard Math (27\% vs.\ 13\%) by substituting elegant derivation with optimized brute-force computational iteration. When the CoT condition forced both models to attempt formal algebraic derivations, GPT-4o collapsed: under CoT, Claude holds Hard Math at 20\% against GPT-4o's 13\% (Figure~\ref{fig:heatmap}). This reversal confirms that GPT-4o's zero-shot strength in CP mathematics relies heavily on computational heuristics rather than formal proofs.

\subsection{Wrong Answer as the Persistent Bottleneck}
\label{sec:discussion:wa}
Despite the introduction of CoT, the most significant empirical signature in our dataset remains the dominance of the Wrong Answer (WA) verdict. Under CoT prompting, WA accounted for 183 of GPT-4o's 199 unaccepted solutions and 80 of Claude's 115. 

This indicates that both models possess a robust grasp of Python syntax and standard API usage, successfully executing code within the strict 5-second limit. The bottleneck is strictly algorithmic correctness. For practitioners building coding assistants, this implies that integrating better code formatters, linters, or compiler-feedback loops will yield marginal returns. Standard prompt engineering (like CoT) fails to bridge this gap; the performance deficit must be addressed at the foundational reasoning layer.

\subsection{Degradation Curves and Capability Ceilings}
\label{sec:discussion:curves}
Comparing the two models across the taxonomy reveals distinct degradation curves. Claude's degradation is back-loaded: it maintains remarkable stability through the Medium tier before collapsing at the Hard tier. A notable exception occurs in Graphs, where Claude maintains a comparatively stable pass rate across the Medium and Hard tiers (53\% and 47\%) by successfully identifying structural graph-theoretic templates (e.g., Strongly Connected Components, 2-SAT) on which GPT-4o scores only 13\%. GPT-4o, by contrast, degrades steeply at every step and bottoms out at 0\% on the hardest DP and Binary Search problems.

\subsection{Limitations}
\label{sec:discussion:limitations}
We acknowledge several limitations in our methodology that restrict the generalizability of these findings:
\begin{enumerate}
    \item \textbf{Instruction Adherence and Token Limits:} Claude frequently struggled to balance verbose CoT generation with the 2048-token limit, occasionally resulting in truncation.
    \item \textbf{Public Test Cases:} Submissions were evaluated against public test cases (pretests) rather than hidden system tests. Because the pretests are limited, the near-absence of Time Limit Exceeded verdicts reflects the lightweight tests as much as the code's asymptotic efficiency; a rigorous evaluation against complete private test suites would likely lower the aggregate pass rates for both models.
    \item \textbf{Language Bias:} All evaluations were conducted in Python 3. Because competitive programming heavily favors C++, the models' exposure to CP concepts during pre-training may be disproportionately weighted toward C++.
    \item \textbf{Sample Variance:} While our stratified sampling ensured category balance, the limit of 15 problems per cell introduces substantial variance; a single problem corresponds to roughly 6.7 percentage points. Cell-level observations (such as the Hard Mathematics inversion, which rests on a difference of two problems) should therefore be read as suggestive rather than statistically confirmed. Expanding the taxonomy grid to a larger $N$ per cell, and reporting confidence intervals, would increase the statistical confidence of the domain-specific anomalies observed.
\end{enumerate}
\section{Conclusion}
\label{sec:conclusion}
In this paper, we introduced a 2D evaluation taxonomy to systematically measure LLM performance on competitive programming tasks across algorithm categories and difficulty tiers. By conducting an ablation study evaluating GPT-4o and Claude Sonnet 4.6 under both Direct and Chain-of-Thought (CoT) prompting conditions, we demonstrated that aggregate pass rates mask severe, domain-specific reasoning deficits. Furthermore, we established that standard prompt engineering paradigms do not cleanly generalize to strict execution-based environments.

\subsection{Summary of Findings}
\label{sec:conclusion:summary}
Our empirical evaluation yielded three primary findings:
\begin{enumerate}
    \item \textbf{The CoT Penalty:} Contrary to standard NLP benchmarks, forcing zero-shot CoT severely penalizes GPT-4o (dropping its aggregate AC rate from 46.0\% to 36.8\%). This is driven by "context poisoning," where the model hallucinates flawed algorithmic proofs (particularly in Greedy logic) that subsequently corrupt its code generation. 
    \item \textbf{Formatting Collapse:} Extended text generation mechanicalizes failure. While Claude Sonnet 4.6 maintained a high logical baseline under CoT (63.5\% AC), the token exhaustion associated with verbose mathematical proofs caused its markdown adherence to break down, more than tripling its Compile Errors (from 9 to 31).
    \item \textbf{The Algorithmic Logic Bottleneck:} The dominance of the Wrong Answer (WA) verdict across both conditions confirms that both models generate syntactically valid Python code that runs within the time limit, but fundamentally fail at mathematical reasoning and edge-case simulation.
\end{enumerate}

\subsection{Future Work}
\label{sec:conclusion:future}
The failure of zero-shot CoT to resolve algorithmic blind spots indicates that single-pass generation has reached a structural reasoning ceiling in competitive programming. Future research must investigate dynamic test-time compute—specifically, whether models can utilize isolated execution sandboxes to verify exchange arguments and recurrence relations prior to final output. Additionally, evaluating multi-agent debate and execution-guided self-repair on our taxonomy will determine if iterative compiler feedback can successfully guide models to correct the deep logical flaws inherent in dynamic programming and greedy failures.
\section*{Acknowledgements}
The authors acknowledge the researchers at Google DeepMind for the open-source release of the CodeContests dataset, as well as the broader Codeforces competitive programming community for providing the robust problem structures that made this evaluation possible.

\bibliographystyle{plainnat}
\bibliography{references}


\newpage
\appendix

\begin{center}
    \LARGE \textbf{Appendices}
\end{center}
\vspace{1em}

\section{Prompt Templates}
\label{app:prompts}
To ensure complete reproducibility, we provide the exact prompt templates used in our evaluation pipeline. The \texttt{\{problem['name']\}} and \texttt{\{problem['description']\}} fields were dynamically populated from the filtered CodeContests dataset during execution.

\subsection{User Prompt (Universal)}
\begin{lstlisting}[basicstyle=\small\ttfamily, breaklines=true]
Problem: {problem['name']}
{problem['description']}

Write a complete Python 3 solution.
\end{lstlisting}

\subsection{System Prompt (Condition A: Direct Generation)}
\begin{lstlisting}[basicstyle=\small\ttfamily, breaklines=true]
You are a competitive programmer. Solve the given problem in Python 3.
Output ONLY the complete Python code inside a markdown block. No explanations.
The code must read from stdin and print to stdout.
\end{lstlisting}

\subsection{System Prompt (Condition B: Chain-of-Thought)}
\begin{lstlisting}[basicstyle=\small\ttfamily, breaklines=true]
You are a competitive programmer. Solve the given problem in Python 3.
First, write a detailed algorithmic plan and mathematical proof of your approach.
Then, output the complete Python code inside a markdown block.
The code must read from stdin and print to stdout.
\end{lstlisting}

\section{Dataset Statistics}
\label{app:dataset}
To construct the balanced evaluation taxonomy of 315 problems, we sampled exactly 15 problems per intersection of Algorithm Category and Difficulty Tier. Table \ref{tab:pool_sizes} details the total available pool of valid, Codeforces-sourced problems within the CodeContests dataset that met our strict filtering criteria before sampling. 

The smallest pool (Binary Search, Easy) contained 72 problems, validating that a strict sample size of 15 problems per cell was methodologically sound without risking duplicate sampling or dataset exhaustion.

\begin{table}[htbp]
  \centering
  \caption{Total available problem pool sizes in the filtered CodeContests dataset before balanced sampling ($n=15$ per cell).}
  \label{tab:pool_sizes}
  \begin{tabular}{lccc}
    \toprule
    \textbf{Algorithm Category} & \textbf{Easy (800--1400)} & \textbf{Medium (1401--1900)} & \textbf{Hard (1901--3500)} \\
    \midrule
    Dynamic Programming & 118 & 440 & 1100 \\
    Graphs              & 80  & 286 & 725 \\
    Greedy              & 547 & 475 & 316 \\
    Binary Search       & 72  & 168 & 224 \\
    Mathematics         & 476 & 305 & 399 \\
    Data Structures     & 99  & 105 & 277 \\
    Implementation      & 631 & 212 & 115 \\
    \bottomrule
  \end{tabular}
\end{table}

\end{document}